\documentclass{PoS}

\title{Using evolutionary algorithms to extract field theory mass spectra}

\ShortTitle{Using evolutionary algorithms to extract field theory mass spectra}

\author{Georg M. von Hippel\thanks{New address: DESY, Platanenallee 6, 15738 Zeuthen, Germany.}, Randy Lewis, and \speaker{Robert G. Petry}\\%
         Department of Physics, University of Regina, Regina, SK, Canada,
         S4S~0A2\\
        E-mail: \email{rob.petry@uregina.ca}}

\abstract{The spectrum of masses from a lattice QCD simulation
  may be found by fitting exponential functions to correlators
  of operators possessing the quantum numbers of the particles
  of interest.  The ability of evolutionary algorithms to find
  globally optimized solutions containing a variable number of
  states across multiple data sets is exploited to provide a
  promising solution to the problem of finding these fits.}

\FullConference{The XXV International Symposium on Lattice Field Theory\\
                 July 30 - August 4 2007\\
                 Regensburg, Germany}

\begin{document}
%%%%%%%%%%%%%%%%%%%%%%%%%%%%%%%%%%%%%%%%%%%%%%%%%%%%%%%%%%%%%%%%%%%%%%%%%%%
\section{Introduction}
\label{sec:introduction}
To calculate the hadronic energy spectrum using lattice QCD the
procedure is in principle straightforward.  First one must identify
quantum numbers of the channel $S$ of interest which may include the
particle's intrinsic angular momentum\footnote{More accurately, since
  one works on the lattice one is interested in the corresponding
  lattice quantum number $\Lambda$ which labels an irreducible
  representation of the octahedral group which corresponds to the
  quantum number~$J$ of the broken continuous rotational symmetry.
  For instance, $J=0 \rightarrow \Lambda=A_1$, $J=1 \rightarrow
  \Lambda=T_1$. Higher $J$ correspond often to multiple octahedral
  irreps.  See, for example,
\cite{Lacock:1996vy,Harnett:2006fp}
for mesons
and
\cite{Basak:2005aq,Basak:2005ir}
for baryons and references therein.}
$J$, parity $P$, charge conjugation $C$, isospin $I$, etc.
Next one constructs the corresponding operators $\Phi^S$ and
$\Phi^{S\dagger}$ which will destroy and create a state with this symmetry.
If one calculates the two-point correlator\footnote{For the sake of
  the discussion assume translationally invariant zero momentum
  operators
\cite{Edwards:2003mv}
.}
\begin{equation}
  G(t)=\left<0\left|T\Phi^S(t)\Phi^{S\dagger}(0)\right|0\right>\;,
\end{equation}
then the energy state spectra must be extracted from the result.
Assuming periodic boundary conditions and a mesonic operator, the fit
function is of the form\footnote{The asymmetries which arise in baryon
  correlation functions require only slight modification of this
  discussion which is restricted for simplicity to meson spectra.}
\begin{equation}
\label{eq:singlecorrfit}
  G(t)=\sum_{n=0}^{n_{max}} Z_n \left(e^{-E_n t}+e^{-E_n (T-t)} \right)\;,
\end{equation}
where here $T$ is the temporal extent of the lattice.
The problem addressed by this paper is how to find the coefficients
and energy states
\begin{equation}
  G=\{(Z_n,E_n): n=1,\ldots,n_{max}\}\;,
\end{equation}
which minimize $\chi^2(G)/n_{dof}(G)$. Here, due to timestep
correlations, we have the correlated $\chi^2$
involving the covariance matrix $\sigma_{ij}$ defined by
\begin{eqnarray}
  \chi^2(G) & = & \sum_{t_i,t_j}( \overline{G_i} - G(t_i))
  (\sigma^{-1})_{ij} (\overline{G_j} - G(t_j))\;, \\
  \sigma_{ij} & = & \overline{G_iG_j}-\overline{G_i}\;\overline{G_j}\;.
\end{eqnarray}
Critically one notes that the number of degrees of freedom,
\begin{equation}
    n_{dof}(G) = (t_{max}-t_{min}+1) - 2 n_{max}\;,
\end{equation}
depends on $n_{max}$, the number of terms in a given fit.
Since the latter is unknown, one has a discontinuous optimization
problem having a solution space spanning multiple dimensions.  We
propose the use of an evolutionary algorithm to solve the problem.  A
complementary discussion of our approach may be found in
\cite{von_Hippel:2007ar}
.
%%%%%%%%%%%%%%%%%%%%%%%%%%%%%%%%%%%%%%%%%%%%%%%%%%%%%%%%%%%%%%%%%%%%%%%%%%%
\section{Evolutionary Algorithms}
\label{sec:ea}
\emph{Evolutionary} or \emph{genetic} algorithms
use the concept of \emph{natural selection} to solve function
optimization problems.  (See 
\cite{von_Hippel:2007ar}
and references therein.)  The terminology reflects this.  Candidate
solutions such as $G(t)$ are \emph {organisms}. The internal encoding
of the solution is its \emph{genotype}, here $G=\{(Z_n,E_n):
n=1,\ldots,n_{max}\}$.  The target function, in our case
$f(G)=-\chi^2(G)/n_{dof}(G)$, is the \emph{fitness} of the
organism.\footnote{We introduce a minus sign in $f(G)$ to ensure a
  higher value indicates a fitter organism.}  Each step in the
algorithm produces a new \emph{generation} $P_\tau$ of individuals.

There are many ways of implementing an evolutionary algorithm. What we
use is representative:
\begin{enumerate}
\item Create the first generation $P_0$ with $N$ randomly generated
  individuals.\footnote{Here $N=(N_{elite}+N_{diversity})^2+N_{mutant}\;$, with the latter constants defined in the algorithm.}
\item Derive $P_{\tau+1}$ from $P_{\tau}$ as follows:
  \begin{enumerate}
  \item \emph{Mutate} each member of the population $P_\tau$ with a
    fixed (small) probability.\footnote{We do not mutate the fittest
      organism to ensure that it will survive to the next generation.}
  \item Select the fittest $N_{elite}$ organisms and a further
    $N_{diversity}$ random organisms placing all their pairwise
    \emph{offspring} into $P_{\tau+1}$.\footnote{Here the offspring of
      the diagonal ``pairs'' between identical organisms is
      considered just a copy of the original organisms themselves
      and we thereby are including the elite in the next generation.}
  \item Add $N_{mutant}$ forced mutations of random elements in the
    elite to $P_{\tau+1}$ to explore the solution space around the
    elite.
  \end{enumerate}
\item Repeat until a suitable termination criterion is
  reached.\footnote{For instance, one may require a minimum number of
    generations be exceeded and that a fixed number of generations
    pass with no improvement in fitness of the best organism. A limit
    on the maximal number of generations may also be imposed.}
\end{enumerate}

In addition to these generic steps one must specify how mutation and
breeding are accomplished within the population.  For the case of a
fit to a single correlator, \emph{mutation} of an individual may
include:
\begin{itemize}
\item Adding or removing a random element $(Z_n,E_n)$ from the genotype's list.
\item Replacing each $(Z_n,E_n)$ by $(Z_n+\Delta Z_n,E_n+\Delta E_n)$
  where $(\Delta Z_n,\Delta E_n)$ are random Gaussian
  deviates.\footnote{Here the standard deviation of the added noise
    can be tuned to the fitness of our genotype $G$ by making it
    proportional to \mbox{$(1-e^{-\alpha\chi^2(G)/n_{dof}(G)})$}
    for some fixed $\alpha$.}
\item Doing a local (e.g. Levenberg-Marquardt) optimization of the
  fit.\footnote{Inclusion of such Newtonian optimizations into
    evolutionary algorithms is often found to be useful
    \cite{Moscato:1989}.
    We restrict the mutation to a fixed number of steps of the local
    optimization for the sake of efficiency.  Greater efficiency might
    be achieved by noting that the linearity of a known fit function
    of the form (\ref{eq:singlecorrfit}) admits local optimization
    with numerical methods which exploit the ability to separate
    linear and non-linear parameters
\cite{Golub:1973:DPI}
.  However such methods will not further aid in the discontinuous
general problem we are solving of finding said function since here
the parameter space is not fixed.}
\end{itemize}
\emph{Breeding} or \emph{crossover} of distinct parent organisms
$par1$ and $par2$ produces two $child$ organisms.  To produce each
child, take corresponding ordered pairs in the parent\footnote{For
  parents having different numbers of ordered pairs we copy the
  extra pairs of the longer parent into the child.} and generate
independent uniformly distributed random numbers
$(x,y)\in[-\delta;1+\delta]$.\footnote{The use of $\delta$ allows for
  the possibility of extrapolation in addition to interpolation
  between the parents' parameters, thereby avoiding an unwanted rapid
  contraction to a central point
\cite{Allanach:2004my}
.} The child element becomes:
\begin{equation}
  (Z^{child},E^{child}) = (xZ^{par1}+(1-x)Z^{par2},yE^{par1}+(1-y)E^{par2})\;.
\end{equation}
Sample fits to single correlators may be found in
\cite{von_Hippel:2007ar}
.
%%%%%%%%%%%%%%%%%%%%%%%%%%%%%%%%%%%%%%%%%%%%%%%%%%%%%%%%%%%%%%%%%%%%%%%%%%% 
\section{Fitting Multiple Correlators}
\label{sec:multiple}
\begin{figure}[h]
  \begin{center}
    \includegraphics[width=6cm,keepaspectratio]{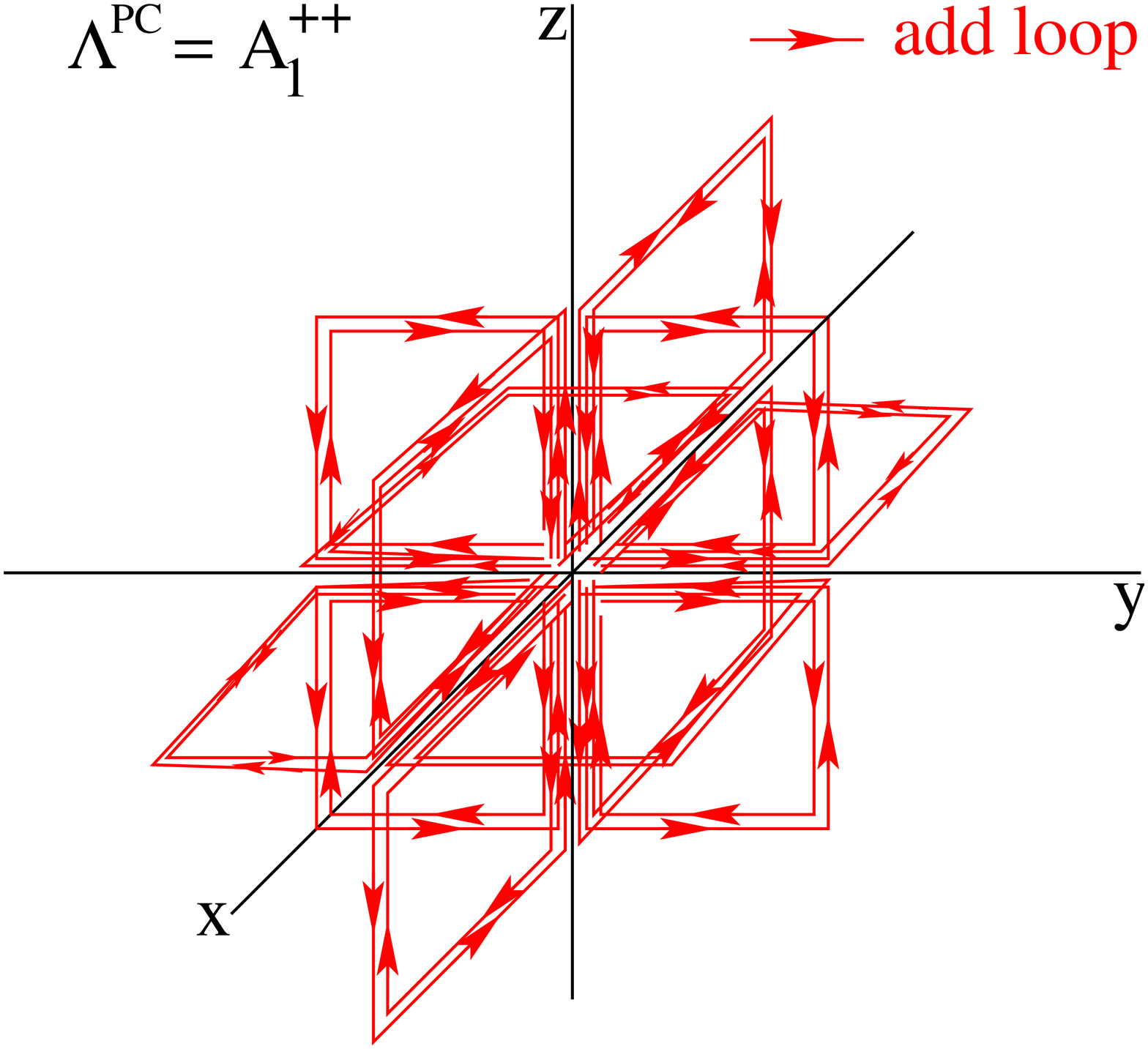}
    \hspace{1cm}
    \includegraphics[width=6cm,keepaspectratio]{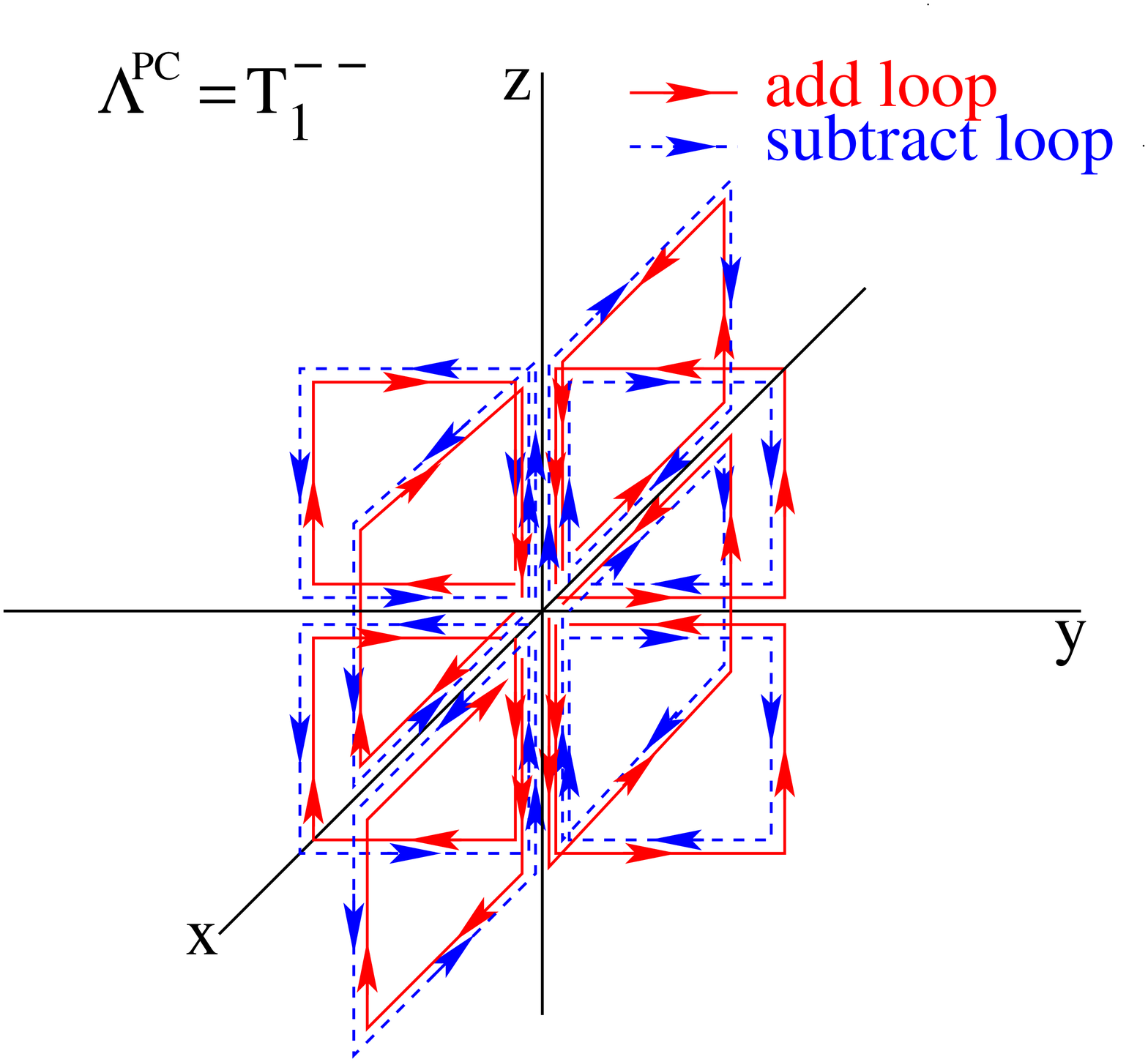}
  \end{center}
  \vspace{-4mm}
  \caption{Extended gauge field structures contributing the
    $\Lambda^{PC}$ shown may be combined via octahedral Clebsch-Gordan
    coefficients to quark operators at the origin to create a greater
    set of operators which couple to desired meson channels.  The
    $T_1$ ``vector'' structure on the right has three components; only
    the one symmetric about the $z$ axis is shown.  See
\cite{Harnett:2006fp}
and references therein.}
  \label{fig:clebsch}
\vspace{-2mm}
\end{figure}

Having shown how to fit a single correlator to extract the spectrum it
contains, we now generalize this to discuss the more practical problem
of fitting several correlators.  Fitting multiple correlators
representing a single channel $S$ is desirable as more data allows the
resolution of a larger number of states with greater accuracy.  One
typically creates \emph{many} operators for the channel $S$ through
group theory methods\footnote{This may be done from the top down by
  creating an operator space and using group theory projection to
  extract operators with quantum numbers $S$ or from the bottom up by
  using Clebsch-Gordan coefficients to construct the desired operators
\cite{Basak:2005aq,Basak:2005ir}
.} or by transforming
existing operators in ways which conserve their channel properties.
An example of the former may be seen in figure~\ref{fig:clebsch},
while quark and link smearing of operators which map $\Phi^{S}
\stackrel{\scriptscriptstyle{\mathrm{smear}}}{\longrightarrow}
\Phi\,'^{S}$ is an example of the latter.
The entire set of operators corresponding to $S$ allows one in
principle to evaluate an entire \emph{correlator matrix} $G_{ij}(t)$
between them:
\begin{equation}
  \{\Phi^S_i: i = 1,\ldots, i_{max}\}\Rightarrow G_{ij}(t)\;,
\end{equation}
all or some subset of which is to be evaluated and fit.

Since the correlator matrix grows as the number of operators squared,
and because the off-diagonal entries have slightly different
functional forms, consider the special case of fitting multiple
diagonal correlators $\{G_{ii}, i=1,\ldots i_{max}\}$ with an
evolutionary algorithm.  This will require changes to our
single-correlator problem.  First we modify the genotype with
$(Z,E)\rightarrow (Z,I)$, where index $I$ points to a state list
$(E_1, \ldots ,E_{m_{max}} )$ common to all the
correlators.\footnote{Here the integer index $I$ is taken modulo
  $m_{max}$ to ensure the coefficient points to an actual energy
  state.}
The full genotype becomes
\begin{eqnarray}
  \label{eq:multidatasetfit}
  \mathrm{Fit\ Genotype}
  & = & ( \mathrm{Dataset\ coefficients},\mathrm{Energy\ state\ list} ) \nonumber \\
  & = & ( (\mathrm{Dataset\ 1\ coefficients},\ldots ),\mathrm{Energy\ state\ list}  ) \nonumber \\
  & = & ((( (Z_1^{(1)},I_1^{(1)}), \ldots, (Z_{n_{max}^{(1)}}^{(1)},I_{n_{max}^{(1)}}^{(1)})), \ldots ), (E_1, \ldots ,E_{m_{max}} ))\;.
\end{eqnarray}
The fitness function $f(G)=-\chi^2(G)/n_{dof}(G)$ is modified due to
having multiple datasets to
\begin{eqnarray}
  \chi^2(G) & = & \sum_{i=1}^{i_{max}}\chi^2(i)\;, \\
  n_{dof}(G) & = & n_{data} - m_{max}-\sum_{i=1}^{i_{max}}n_{max}^{(i)}\;,
\end{eqnarray}
where $n_{data}$ is the product of the number of timesteps fit and the
number of correlators.
The complexity of the genotype permits enhanced evolutionary
operations.  As a nested hierarchy of lists,
(\ref{eq:multidatasetfit}) admits more complicated list-based
mutations and breeding.  Integer indices may be bred and mutated
bitwise.  Finally a reduction mutation which orders masses and
coefficients is useful to encourage the algorithm to converge to a
single representation of the solution.

\begin{figure}[p]
\begin{center}
\includegraphics[width=.75\textwidth,keepaspectratio]{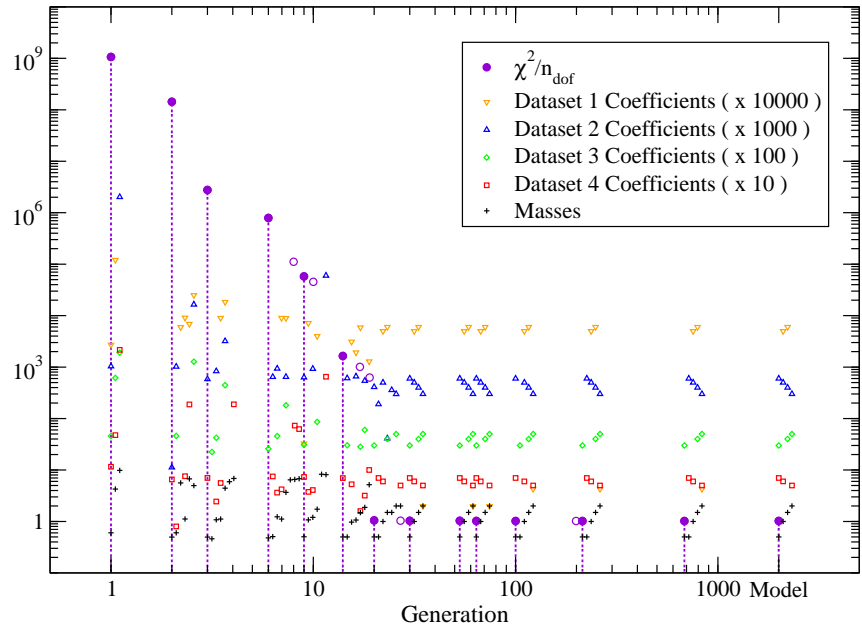}
\end{center}
\caption{Masses and coefficients of simultaneous fit to four synthetic correlators.}
\label{fig:synthdata}
\end{figure}

\begin{figure}[hp]
  \begin{center}
    \includegraphics[width=.75\textwidth,keepaspectratio]{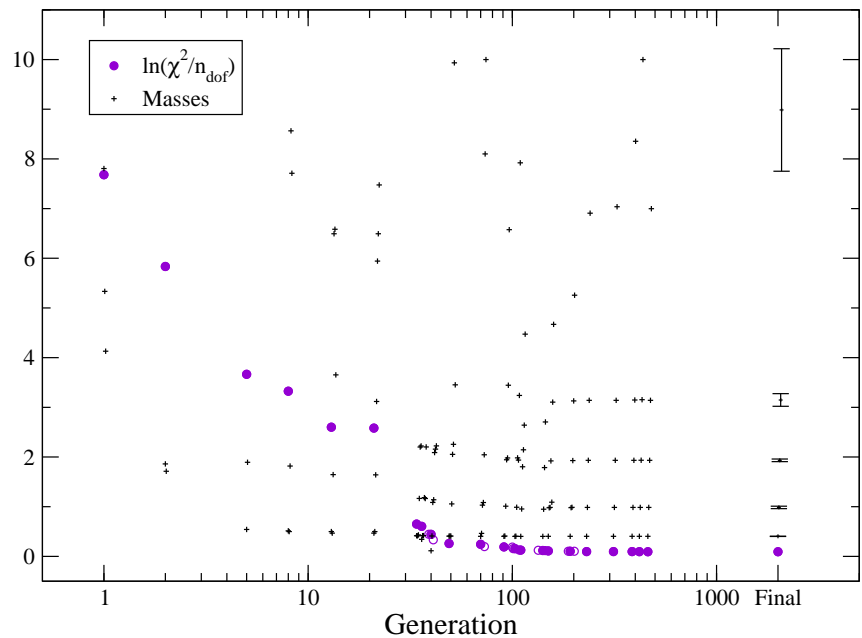}
  \end{center}
\caption{Masses of simultaneous fit to eight $\rho$ meson ($T_1^{--}$) correlators.
}
\label{fig:multirho}
\end{figure}

In figure~\ref{fig:synthdata} the effectiveness of the algorithm to
find a known solution is shown.  Four synthetic correlators, each with
$48$ timesteps, were created by adding noise to the model function
depicted on the right of the plot.  There one sees four masses
displaced horizontally with the corresponding coefficients in each
respective dataset plotted vertically above the corresponding mass.
The left side of the plot shows the best fit of each generation.  The
plot also displays $\chi^2/n_{dof}$ of the best fit (circles) and one
sees it converge to 1 as expected as the fit improves.\footnote{See
\cite{von_Hippel:2007ar}
for further discussion of this plot.} A simultaneous fit to actual
data of eight diagonal $\rho$ meson (i.e. $\Lambda^{PC}=T_1^{--}$)
correlators\footnote{Simulation details: Wilson quarks, $\beta=6.0$,
  $\kappa=.1554$, $20^3\times 48$, $600$ configurations, quenched.}
is shown in figure~\ref{fig:multirho}.  Only the energy states are
shown of the best fit of each generation up to generation 600.  The
coefficients in each dataset, a further $30$ parameters in the final
fit, are not shown.  The last column depicts the best fit found with
bootstrap errors produced via Levenberg-Marquardt fits to bootstrap
configurations with its fixed functional form.\footnote{Note that as
  well a Levenberg-Marquardt optimization was done on the best fit
  found to produce the final result.}

%%%%%%%%%%%%%%%%%%%%%%%%%%%%%%%%%%%%%%%%%%%%%%%%%%%%%%%%%%%%%%%%%%%%%%%%%%%
\section{Advantages of Evolutionary Algorithm Fitting}
\label{sec:advantages}
To conclude, we present advantages of the evolutionary algorithm
fitting method. For one it is a global optimization method which is
furthermore independent of initial conditions. The solution space is
discontinuous as it spans multiple dimensions, arising from the fact
that one does not know the exact functional form \emph{a priori}.  The
evolutionary algorithm fitting method is capable of handling this
problem; by minimizing $\chi^2/n_{dof}$ it finds the number of states
in the data in a natural manner.  As well, the ability to identify
whether a state exists or not in an individual correlator in a
discrete way means that evolutionary fitting can in principle identify
which orthogonal irreps~$\Lambda$ a state straddles and hence aids in
the identification of its continuous angular momentum~$J$
\cite{Harnett:2006fp}
.

Computationally, evolutionary algorithms are inherently
parallelizable.  One can break populations into \emph{islands}
breeding on different nodes/CPUs largely independently with only
occasional migration between them.  Large datasets can also be
partitioned with sub-genotypes being initially evaluated and then
stitched together for further evaluation on the entire dataset.
Evolutionary fitting does not require evaluation of the full
correlator matrix, which allows for inexpensive asymmetrical smearing
between the sink and source operators.\footnote{Indeed, our algorithm
  permits consideration of additional correlators which would not even
  be allowed in a correlator matrix, namely those for which only a
  single operator at the sink or source has the definite quantum
  numbers of channel~$S$.  The other operator could potentially couple
  to many channels beyond that of interest.}  This also means one can
restrict oneself to evaluating only the diagonal correlators of the
correlator matrix where one expects to have the strongest signals.
This in turn allows a wider assortment of operators to be evaluated.
Finally, there is the potential for combination with established
methods.\footnote{One can determine priors for Bayesian fits using
  distributions of parameters spread across population islands
\cite{von_Hippel:2007ar}.
In the variational method, one can diagonalize the operators in the
usual way and then fit only the new diagonal correlators with our
fitting algorithm.  This would maximize the information in a minimized
amount of data for the subsequent evolutionary fit. See
\cite{Burch:2006dg}, for example, on the variational method used with
meson correlators.}  We are currently applying this evolutionary
fitting method to analyze all the operators detailed in
\cite{Harnett:2006fp}
with promising results.

%%%%%%%%%%%%%%%%%%%%%%%%%%%%%%%%%%%%%%%%%%%%%%%%%%%%%%%%%%%%%%%%%%%%%%%%%%%
\vspace{-1mm}
\section*{Acknowledgements}
\vspace{-1mm}
We thank George T. Fleming for drawing our attention to reference
\cite{Golub:1973:DPI},
and Richard M. Woloshyn for providing gauge field configurations and
propagators
\mbox{\cite{AbdelRehim:2005gz}}.
This work was supported in part by the Natural Sciences and
Engineering Research Council of Canada, the Canada Foundation for
Innovation, the Canada Research Chairs Program and the Government of
Saskatchewan.

\end{document}